\documentstyle[prl,aps,epsf]{revtex}

\newcommand{\A}{{\mathbf A}}
\newcommand{\D}{{\mathbf D}}
\newcommand{\k}{{\mathbf k}}
\newcommand{\K}{{\mathbf K}}
\newcommand{\muhat}{\hat{\mu}}
\newcommand{\Upsy}{\Upsilon_\parallel}
\newcommand{\Ly}{L_\parallel}

\begin{document}
\advance\textheight by 0.2in
\draft
\twocolumn[\hsize\textwidth\columnwidth\hsize\csname@twocolumnfalse\endcsname

\title{High-$T_c$ Superconductors in Applied Magnetic Fields Parallel
  to the CuO Planes:\\ First Order Transition with Slow Onset of Resistivity}

\author{Peter Olsson and Petter Holme}

\address{Department of Theoretical Physics, Ume{\aa} University, 
  901 87 Ume{\aa}, Sweden}

\date{\today}   

\maketitle

\begin{abstract}
  The three dimensional uniformly frustrated XY model is used as a
  model of a high temperature superconductor in an applied magnetic
  field parallel to the CuO-planes.  Through Monte Carlo simulations
  with anisotropy $\eta^2 = 10$ on large lattices we find evidence for
  a first order transition.  Earlier simulations and theoretical
  treatments are discussed and the experimentally found smooth onset
  of resistivity is suggested to be due to a large potential barrier
  against vortex line motion above $T_c$ present for perfect alignment
  of the applied field.
\end{abstract}

]

%\paragraph{Introduction}
The behavior of cuprate superconductors in applied magnetic fields has
been, and continue to be, a very active area of research. The most
studied geometry is with the applied field perpendicular to the
CuO-planes. In that case, for clean samples, a first order transition
associated with the melting of the Abrikosov lattice is by now well
established both through calorimetric measurements\cite{Calorimetric}
and measurements of the jump in the magnetic induction\cite{Magnetic}.
Also, a number of transport measurements are consistent with a first
order transition.

The geometry with the applied field \emph{parallel} to the planes, has
not been that thoroughly examined.  It has, however, been found that
the sharp increase in resistivity associated with the first order
melting transition vanishes for perfect alignment of the field, and
that the temperature-dependence of the resistivity is consistent with
a continuous transition\cite{Kwok-H.ab}.  This remarkable finding
naturally rises the question of the possibility of a continuous vortex
line lattice melting, and the commonly accepted theoretical picture is
that the system undergoes a continuous transition analogous to the
nematic to smectic transformation in liquid
crystals\cite{Balents_Nelson:95}.  A more recent
experiment\cite{Grigera_MONC} suggests that the transition from a
vortex liquid to a vortex solid phase takes place in two steps, with
an intermediate `possible smectic' phase.

In the first Monte Carlo study of this model\cite{Hu_Tachiki} the
authors examined a model with anisotropy $\eta^2 = 100$ (see below),
and argue that the behavior of both the helicity modulus and the
specific heat suggest a continuous transition.  However, a comparison
between their Fig.~2 and corresponding data for a 2D XY model of the
same size shows a remarkable similarity, suggesting that it is the
individual layers that dominate the behavior of this three-dimensional
system. This makes it difficult to draw any conclusions based on these
results.

In this Letter we report on a study of a similar system, but with
anisotropy $\eta^2 = 10$. The smaller value for the anisotropy is
chosen with the hope to get a more clearly visible signal from the
transition.  The behavior of the system is clearly dependent on the
system size.  Only for very large systems does it become clear that
the melting transition is first order instead of continuous.  We also
discuss problems with the suggested smectic phase and suggest a simple
mechanism behind the apparently continuous onset of resistivity at the
transition.  The suggested scenario is directly open for experimental
verification.

%\paragraph{Definition of the model}
For the study of superconductors in applied magnetic fields we make
use of the uniformly frustrated anisotropic 3D XY model, with the
Hamiltonian
\begin{equation}
  {\cal H} = - \sum_{i,\muhat} J_\mu \cos(\theta_{i + \muhat} -
  \theta_i - A_{i\mu}),
  \label{Hamiltonian}
\end{equation}
where the sums are over the lattice points $i$ and $\muhat = \hat{x}$,
$\hat{y}$, and $\hat{z}$.  To include a magnetic field in the
$y$-direction we choose $A_{i\mu} =
(2\pi/\phi_0)\int_i^{i+\muhat}\A\cdot d\mathbf{l}$ such that
\begin{displaymath}
  \frac{1}{2\pi}\D \times A_{i\mu} = f\hat{y},
\end{displaymath}
where $\D$ is the discrete difference operator. We use equal couplings
in the $x$-$y$ planes, $J_x = J_y = J_{xy}$, but a different coupling,
$J_z$, between the planes. The anisotropy is given by the parameter
\begin{equation}
  \eta = \sqrt{\frac{J_{xy}}{J_z}} = \frac{\lambda_z}{\lambda_{xy}}
  \frac{d}{\xi_{xy}},
  \label{eta}
\end{equation}
where the relation to the physical parameters, the penetration lengths
$\lambda_\mu$ in the different directions, the inter-plane distance $d$
and the bare vortex core size in the planes $\xi_{xy}$, is given by
the second equality. A derivation and justification of Eqs.\ 
(\ref{Hamiltonian}) and (\ref{eta}) may be found in Ref.\ 
\cite{Chen_Teitel:97}.

To examine the behavior of this system we have simulated Eq.\ 
(\ref{Hamiltonian}) with periodic boundary conditions, $f = 1/24$, and
several values for the anisotropy and system sizes.  The results
reported here are for $\eta^2=10$ and $L_x = L_z = 48$, and $\Ly =
48$, 96, 192, and 384.  We measure several different quantities. The
superconducting coherence in the different directions as examined
through the helicity moduli $\Upsilon_x$, $\Upsilon_z$, and $\Upsy
\equiv \Upsilon_y$\cite{Chen_Teitel:97}.  To study the melting of the
vortex lattice we measure the structure factor $S(\k_\perp)$ where
$\k_\perp$ is a wave vector perpendicular to the field and monitor
$\Delta S = |S(\K) - S(R_x[\K])|$ \cite{Chen_Teitel:97} where $\K$ is
a reciprocal lattice vector of the ordered vortex lattice, and $R_x$
reflects $\K$ through the $x$ axis, and we average over the three
smallest non-zero values of $\K$.  Our Monte Carlo simulations
consist, for the larger systems, typically of $10^5$ passes through
the lattice for equilibration followed by $1\times 10^6$ through
$4\times 10^6$ passes for calculating averages. The runs on the
smaller systems were typically somewhat shorter.

%\paragraph{Melting of the vortex lattice}
For the geometry with the field perpendicular to the CuO layers it has
been found that the vortex lattice melting transition is best examined
by starting with a disordered system and slowly lowering the
temperature until the system finds an ordered configuration with a
vortex lattice\cite{Hu_M_Tachiki}.  However it turns out that the
choice of $\Ly$ -- the extension of the system in the field direction
-- is crucial in order to succeed with this cooling into an ordered
state and obtain the correct transition behavior. With a too large
value of $\Ly$ it becomes very difficult -- or due to limited computer
resources even impossible -- to cool into a lattice. With a too small
$\Ly$ the simulations incorrectly give at hand that there are two
separate transitions with the vortex lattice melting taking place at
the lower temperature followed by the vanishing of $\Upsy$ at a higher
temperature.  The proper choice of the length $\Ly$ depends on the
parameters of the model, $f$ and $\eta$.  In the first simulations
with clear evidence of a first order transition, with $\eta^2 = 10$
and $f = 1/25$, this length was chosen as $\Ly=40$\cite{Hu_M_Tachiki},
but for isotropic systems, $\eta^2 = 1$ and $f=1/20$, the evidence of
a sharp transition was only found for $\Ly =
128$\cite{Olsson_Teitel:xy3f} or $\Ly=120$\cite{Nguyen_Sudbo:98}.

\begin{figure}
  \epsfxsize=8.5truecm          %  \epsfbox{fig-melt.ps}
  \epsfbox{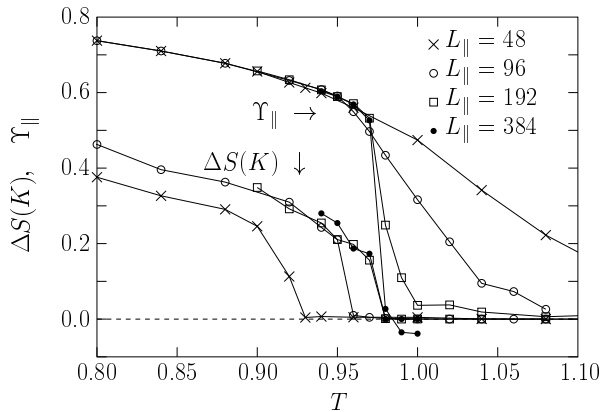}
  \caption{The melting of the ground state vortex lattice for $L_x=L_z
    = 48$ and several different values for $\Ly$. Shown is the
    structure factor together with $\Upsy$. Only for $\Ly\geq192$ does
    it become clear that there is a single transition with
    simultaneous vanishing of both $\Delta S(K)$ and $\Upsy$. }
  \label{fig-melt}
\end{figure}

Guided by the experience with fields perpendicular to the CuO planes
we therefore first tried slowly cooling the system, but we never
succeeded cooling into a lattice.  We therefore instead started the
simulations with an artificially prepared initial configuration with a
perfect vortex lattice and slowly increased the temperature. In this
way we obtained results similar to Fig.~1 of Ref.\ 
\cite{Olsson_Teitel:xy3f}.  The vortex lattice melting as probed by
$\Delta S(\K)$ is shown together with $\Upsy$ in Fig.\ \ref{fig-melt}
for $\Ly = 48$, 96, 192, and 384.

This data shows that one needs at least $\Ly\approx192$ for obtaining
evidence that the vanishing of $\Upsy$ goes together with the melting
of the vortex lattice.  The similarity with data from Ref.\ 
\cite{Olsson_Teitel:xy3f} for the field perpendicular to the planes
strongly suggests a first order transition also for this case with the
field parallel to the planes.

From these simulations for $\eta^2 = 10$ we can of course not
immediately draw any conclusions regarding the character of the
transition for larger $\eta$.  Still, we will argue below that both
the simulation results presented in Ref.\ \cite{Hu_Tachiki} and the
experimentally found smooth onset of the resistivity\cite{Kwok-H.ab}
are consistent with first order transitions.  Since we also find
problems with the continuous melting scenario, we suggest that the
transition actually is first order for all finite values of the
anisotropy.

%\paragraph{Earlier simulations}
Turning to the simulation results for $\eta^2=100$\cite{Hu_Tachiki},
we first note that the choice $\Ly = 40$, or 80, in the light of Fig.\ 
\ref{fig-melt}, not is large enough for giving the correct behavior.
But it also seems that their results are dominated by the behavior of
the (weakly coupled) two-dimensional layers in the $x$-$y$ plane.  To
understand the relation to the 2D XY model we note that it is possible
to choose a gauge with $A_{ix} = A_{iy} = 0$, and the frustration
included in the $A_{iz}$. Considered in this way, the system consists
of some stacked unfrustrated 2D XY planes with a weak interplane
coupling that at some points between the planes is ferromagnetic
($A_{iz} = 0$), at some points is antiferromagnetic ($A_{iz} = \pi$),
and otherwise is something in between.

Fig.~2 of Ref.\ \cite{Hu_Tachiki} shows that the specific heat in the
three dimensional model has a peak around $T = 1.05$. The maximum
value of the plotted quantity is $24\times C \approx 36$. The result
was found to be about the same for all three different system sizes in
the figure, of which the smallest was $48\times 40\times 48$. To
compare with the specific heat for a single 2D layer, we have made
simulations with $L=44$ which is chosen to have close to the same area
as $48\times 40$.  Recall that the peak of the specific heat in the 2D
model is a non-singular feature well separated from the
Kosterlitz-Thouless temperature\cite{Kosterlitz_Thouless} $T_{\rm KT}
\approx 0.892$\cite{Olsson:Kost-fit}.

\begin{figure}
  \epsfxsize=8.5truecm          %  \epsfbox{fig-2D-cv.ps}
  \epsfbox{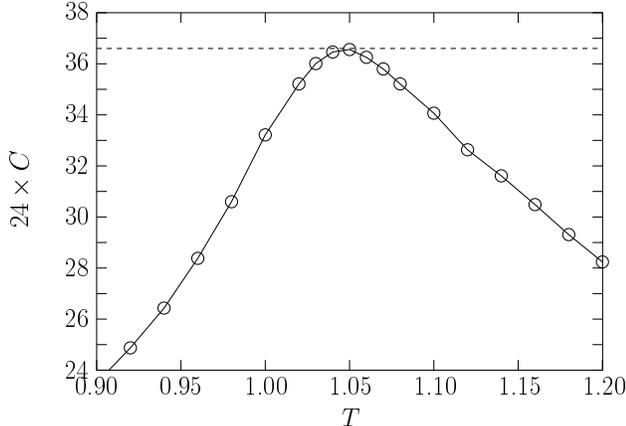}
  \caption{Specific heat for a two-dimensional $XY$ model with $L =
    44$. The normalization is chosen to facilitate direct comparison
    with Fig.~2 of Ref.\ \protect\cite{Hu_Tachiki}. The conclusion
    is that their data is dominated by 2D fluctuations and not should
    be considered as evidence for a continuous transition in the 3D
    system.}
  \label{fig-2D-cv}
\end{figure}

Our results for $24\times C$ for the 2D system are shown in Fig.\ 
\ref{fig-2D-cv}. The similarity to the data of Hu and Tachiki (inset
of their Fig.~2) is striking. Not only the shape of the curves is the
same but also the position and the height of the maximum: at $T
\approx 1.05$ we find the value $\approx 36.6$ which is the same as --
or possibly slightly higher than -- the data in their Fig.~2.  We
therefore conclude that the data for $\eta^2 = 100$ of Ref.\ 
\cite{Hu_Tachiki} is dominated by 2D fluctuations. The absence of any
sharp features in this data should therefore not be taken as evidence
for a continuous transition in the 3D model.

Why is the interplane coupling not more important?  An obvious but
rough way to assess the importance of the coupling between two planes
is to compare the total coupling with the temperature.  The size of
the planes should then be large enough if $L_x \times L_y \times J_z
\gg T$, which for our present values becomes $19.2 \gg 1.05$. This
would lead to the conclusion that the planes are large enough to make
the interplane coupling very significant. The reason why this approach
fails is the presence of the frustrating vector potential $A_{iz}$
that varies between 0 and $2\pi$. This means that the interplane
couplings tend to cancel each other out, and we believe this to be the
reason for the very 2D-like behavior found in the model.

%\paragraph{Continuous melting?}
In the theory by Balents and Nelson\cite{Balents_Nelson:95} the
transition takes place in two steps with an intermediate smectic
phase.  The smectic phase is characterized by order in one direction
only, which results in only certain layers being occupied with vortex
lines and the others being empty.  However, to us this picture doesn't
seem to be a convincing explanation of the experimental findings of
Ref.\ \cite{Kwok-H.ab} for two reasons.

First, the experimentally found smooth onset of resistivity was only
found for very small angles, $\theta< 0.5^\circ$, but there seems to
be no reason to expect that the transition between the smectic and the
liquid phases should be very sensitive to the precise angle of the
applied magnetic field.  The transition between the tilted smectic
phase and the liquid is also expected to be a continuous transition,
which only changes character to first order for large
tilts\cite{Balents_Nelson:95}.
%The extreme sensitivity to the direction of the magnetic field doesn't
%seem to be accounted for in the theory of Ref.\ 
%\cite{Balents_Nelson:95}.

\begin{figure}
  \epsfxsize=8.5truecm          %  \epsfbox{fig-unstable.ps}
  \epsfbox{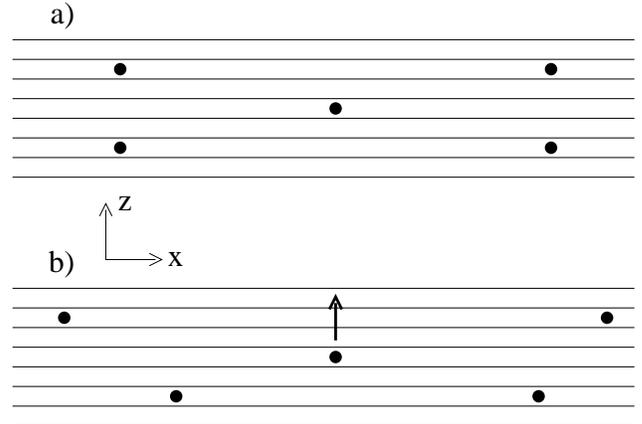}
  \caption{The figures show a side-view of a layered superconductor in
    the direction of the applied field.  Panel a) is a vortex lattice
    and panel b) illustrates a smectic phase which is ordered in the
    $\hat{z}$ direction (only certain layers occupied) but disordered
    in the $\hat{x}$ direction. This disorder may, however, give rise
    to forces (as indicated by the arrow) that tend to move the vortex
    lines from the occupied layers and destroy the smectic order.}
  \label{fig-unstable}
\end{figure}

Second, considering the thermal fluctuations of the vortex lattice,
Balents and Nelson found that the average relative displacements in
the $x$ and $z$ directions are unequal with $\left<u^2_x\right>/a_x^2
\gg \left<u^2_z\right>/ a_z^2$.  This was then taken to suggest that
the Lindemann criterion for melting can be satisfied in the one
direction only and not in the other, leading to a partial melting of
the lattice.  However, this line of reasoning presumes that
$\left<u^2_z\right>$ not is strongly affected by the melting in the
$\hat{x}$ direction, but there is no \emph{a priori} reason for this
assumption.  Figure \ref{fig-unstable} illustrates that the disorder
in the $\hat{x}$ direction may well give rise to forces that destroy
the layered structure.  This suggests problems with the stability of
the smectic phase, which would seem to altogether invalidate the
continuous melting scenario.

%\paragraph{Smooth onset of resistivity}
So far we have both discussed the validity of the claim of a
continuous transition from simulations and some problems with the
theory of continuous melting.  We now turn to the experimentally found
smooth onset of the resistivity\cite{Kwok-H.ab} suggestive of a
continuous transition. We will argue that this observation not is
inconsistent with a first order melting transition.

A current through the system perpendicular to the applied field gives
rise to a force on the vortex lines perpendicular to both the current
and the field.  If the vortex lines move in response to this force the
system dissipates energy, which is experimentally seen as a
resistivity in the sample.  In the geometry of Ref.\ \cite{Kwok-H.ab}
with the current in the CuO planes and the applied field also parallel
to these planes, dissipation is related to the motion of the vortex
lines in the $\hat{z}$ direction.

\begin{figure}
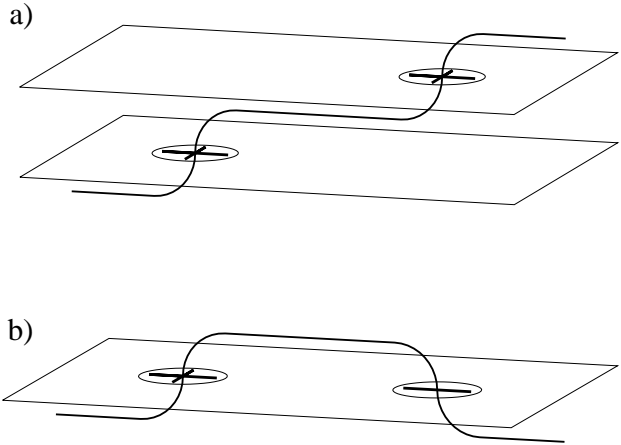

  \epsfxsize=8.5truecm          %  \epsfbox{fig-mech-ne0.ps}
  \epsfbox{figure4a.ps}
  \epsfxsize=8.5truecm          %  \epsfbox{fig-mechanism.ps}
  \epsfbox{figure4b.ps}
  \caption{Motion of vortices and vortex lines. These figures illustrate
    the motion of vortex lines for a) a tilted magnetic field and b)
    perfect alignment.  For perfect alignment of the field there is a
    vortex anti-vortex pair in the CuO plane, and together with the
    motion of an increasingly larger portion of the vortex line to the
    upper layer goes a separation of the vortices in the CuO plane.}
  \label{fig-mechanism}
\end{figure}

An interesting observation is that the mechanism for motion of vortex
lines in the $\hat{z}$ direction becomes qualitatively different for
perfect alignment of the field with the CuO layers as compared to a
tilted magnetic field. This is illustrated in Fig.\ 
\ref{fig-mechanism}. Panel a) shows the situation in the presence of a
tilted magnetic field where the upward motion of a vortex line is
linked to the vortices moving to the left in the figure.  Panel b)
shows the qualitatively different situation for fields perfectly
aligned with the CuO planes. In this case the motion of a vortex line
has to start with a `kink' that then grows bigger until the whole line
has crossed the plane.  The crucial thing to note is that such a kink
gives rise to a vortex anti-vortex pair in the $x$-$y$ plane, and that
this vortex pair separates as the vortex line gradually moves upward
across the plane.

The energy barrier to be overcome for motion of the vortex lines in
Fig.\ \ref{fig-mechanism}a is the pinning of individual vortices. In
contrast, the energy barrier in Fig.\ \ref{fig-mechanism}b is the kink
energy barrier -- related to the vortex anti-vortex interaction in the
plane -- which may be quite large just above $T_c$.  If we assume that
the transition implies a change of this potential barrier from
infinite to quite large, the first order transition should not be
clearly seen in the resistivity.  In a 2D system this potential
barrier diverges as $T\rightarrow T_{\rm KT}$, and as we expect
$T_c\rightarrow T_{\rm KT}$ as the anisotropy increases, and that the
behavior is 2D-like above $T_c$, we believe that this assumption of a
large potential barrier right above $T_c$ is a reasonable one at least
for fairly large anisotropies.

With this mechanism for the transition the very strong dependence of
the resistivity on the angle of the applied field\cite{Kwok-H.ab}
follows naturally.  Perfect alignment of the field corresponds to
two-dimensional XY layers with only thermal vortex pairs and no free
vortices in the low-temperature phase.  But even a small non-zero
angle of the field gives some additional free vortices, and the
presence of them would lead to a qualitatively different behavior.

%\paragraph{Experimental test}
The existence of a first order melting transition should be possible
to verify from measurements of e.g.\ the specific heat.  This would be
a direct experimental test of our proposed picture, since in the
alternative scenario with a continuous transition into a smectic phase
there should be no sharp anomaly in the specific heat.  However, so
far the experiments are inconclusive.  In Ref.\ 
\cite{Schilling_FPWKC:97} the authors found features in the specific
heat both for field parallel and perpendicular to the planes. The work
was later extended to several different values of the angle of the
field, including a parallel field, and it was found that the sizes of
these features scaled with the angle in accordance with a simple
scaling relation\cite{Schilling_FPWKC:98}.  This seems to point to the
existence of a first order melting transition, but not conclusively.
As the authors point out\cite{Schilling_FPWKC:97,Schilling_FPWKC:98},
the reason for the apparent first order character could also be that
the angular resolution was not sufficient to access the true perfect
alignment region $|\theta| < 0.5^\circ$\cite{Kwok-H.ab}.  A carefully
designed experiment should make it possible to discriminate between
the two different proposed pictures.

%\paragraph{Conclusion}
To conclude, we have found a single first order transition in the
anisotropic 3D $XY$ model with magnetic field applied parallel to the
planes.  The finding that very large system sizes are needed for
obtaining the true behavior make us doubt the continuous transitions
reported from simulations on smaller systems.  Some weaknesses with
the commonly accepted theory of continuous vortex lattice melting have
been discussed and we suggest that the reason for the smooth onset of
resistivity is a large potential barrier right above $T_c$ associated
with the separation of vortex anti-vortex pairs which is linked to the
motion of the vortex lines in the $\hat{z}$ direction.

We acknowledge discussions with Professor Steve Teitel and Professor
Petter Minnhagen.  This work has been supported by the Swedish Natural
Science Research Council through Contract No.\ E-EG 10376-312, and by
the resources of the Swedish High Performance Computing Center North
(HPC2N).

\end{document}